\documentclass[aps,twocolumn,showpacs,superscriptaddress,nofootinbib]{revtex4}

\usepackage{graphicx}
\usepackage{dcolumn}

\usepackage{amsbsy}

\newcommand{\qq}{\langle \bar{q}q \rangle}
\newcommand{\sss}{\langle \bar{s}s \rangle}
\newcommand{\uu}{\langle \bar{u}u \rangle}
\newcommand{\dd}{\langle \bar{d}d \rangle}
\newcommand{\qGq}{\langle \bar{q} g_s{\sigma}\cdot {G}q\rangle}
\newcommand{\uGu}{\langle \bar{u} g_s{\sigma}\cdot {G}u\rangle}
\newcommand{\dGd}{\langle \bar{d} g_s{\sigma}\cdot {G}d\rangle}
\newcommand{\sGs}{\langle \bar{s} g_s{\sigma}\cdot {G}s\rangle}
\newcommand{\GG}{\langle \alpha_s \pi^{-1} G^2 \rangle}

\def\aJLone<#1,#2>{#1}
\def\aJLtwo<#1,#2,#3>{#2}
\def\aJLyear<#1,#2,#3,#4>{#3}
\def\aJLpage<#1,#2,#3,#4>{#4}

\def\aJpage<#1,#2,#3>{#3}

\begin{document}

\title{Mixings of 4-quark components in light non-singlet scalar mesons in QCD sum rules}

\author{J.~Sugiyama}%
\email[e-mail: ]{sugiyama@th.phys.titech.ac.jp}%
\affiliation{%
Department of Physics, H-27, Tokyo Institute of Technology,\\ Meguro, Tokyo 152-8551, Japan
}%
\author{T.~Nakamura}%
\email[e-mail: ]{nakamura@th.phys.titech.ac.jp}%
\affiliation{%
Department of Physics, H-27, Tokyo Institute of Technology,\\ Meguro, Tokyo 152-8551, Japan
}%
\author{N.~Ishii}%
\email[e-mail; ]{ishii@ribf.riken.jp}%
\affiliation{Center for Computational Sciences, University of Tsukuba,\\ Tsukuba, Ibaraki 305-8577, Japan %
}%
\author{T.~Nishikawa}%
\email[e-mail: ]{nishi@th.phys.titech.ac.jp}%
\affiliation{%
Department of Physics, H-27, Tokyo Institute of Technology,\\ Meguro, Tokyo 152-8551, Japan
}%
\author{M.~Oka}%
\email[e-mail: ]{oka@th.phys.titech.ac.jp}%
\affiliation{%
Department of Physics, H-27, Tokyo Institute of Technology,\\ Meguro, Tokyo 152-8551, Japan
}%
\date{\today}

\begin{abstract}
Mixings of 4-quark components in the non-singlet scalar mesons are studied in the QCD sum rules.
We propose a formulation to evaluate the cross correlators of $q\bar q$ and $qq\bar q \bar q$ operators 
and to define the mixings of different Fock states in the sum rule.
It is applied to the non-singlet scalar mesons, $a_0$ and $K_0^\ast$.
It is found that the 4-quark operators predict lower masses than the $q\bar q$ operators 
and that the 4-quark states occupy about $70$-$90\%$ of the lowest mass states.
\end{abstract}
\maketitle
\section{introduction}
Hadrons which are not represented by minimal $q\bar{q}$ (meson) or $qqq$ (baryon) configurations 
are called exotics,
and $\Theta^+$ with baryon number $B=1$ and strangeness $S=+1$ is the most prominent example.
Recent developments in hadron spectroscopy are largely enhanced by the discovery of the pentaquark $\Theta^+$~\cite{Nakano}.
The LEPS group, who first observed a peak in the $NK$ spectrum produced by the SPring-8 photon beam, 
has confirmed the resonance in their succeeding experiment~\cite{Nakano2},
while many others have been unsuccessful in producing the resonance in various types of experiments~\cite{Schumacher}.
%
While experimental efforts are obviously needed now,
a large number of theoretical studies have been carried out to explain the mass and the width of $\Theta^+$,
as well as its structure and reactions.
At the same time, various new types of exotic hadrons have been proposed and examined.
In fact, the spectroscopy of old (ordinary) hadrons has been reexamined,
and various possibilities of multiquark components of hadrons have been pointed out.
Among them, the possible 4-quark structures of the scalar mesons are not new,
but rather an old idea~\cite{Jaffe4q,Black}.
The (non-relativistic) quark model based on $SU(6)\times O(3)$ symmetry 
does not easily account for the mass pattern of the lowest-lying scalar-meson nonet, 
($\sigma(600), a_0(980), f_0(980), K_0^\ast(800)$).
($K_0^\ast$ has been indicated in $K\pi$ final states in $J/\psi$ and $D$-meson decays, but has not 
yet been established~\cite{PDG}.)
The expected mass-ordering as $q\bar q$ states would be $m(\sigma) \sim m(a_0) < m(f_0)$,
if we assume ideal mixing, \emph{i.e.}, $\sigma\sim \frac{u\bar u + d\bar d}{\sqrt{2}}$, 
$a_0\sim \frac{u\bar u - d\bar d}{\sqrt{2}} $ and $f_0 \sim s\bar s$.
This pattern does not agree with the one from experiment.
Furthermore, while the scalar mesons are classified as $^3P_0$ states, 
their spin-orbit partners $J= 1$ and $2$ states are not observed in their vicinity.

A possible solution to this puzzle is to consider 4-quark exotic states for the scalar mesons~\cite{Jaffe4q}.
Suppose that diquarks with flavor 3, color 3 and spin 0, \emph{i.e.},
\begin{eqnarray}
&& U=(\bar d\bar s)_{S=0,C=3,f=3},\qquad D=(\bar s\bar u)_{S=0,C=3,f=3},\nonumber\\
&& S=(\bar u\bar d)_{S=0,C=3,f=3}, 
\label{diquark}
\end{eqnarray}
are the building blocks of the scalar mesons.
Then the scalar nonet in the ideal mixing may appear as
\begin{eqnarray}
&& \sigma\sim S\bar S \sim (ud)(\bar u\bar d) \nonumber,\\
&& a_0\sim \frac{1}{\sqrt{2}} (U\bar U - D\bar D) \sim \frac{1}{\sqrt{2}} \left[(ds)(\bar d \bar s)-(su)(\bar s\bar u)\right],
\nonumber\\
&& f_0 \sim \frac{1}{\sqrt{2}} (U\bar U + D\bar D) \sim \frac{1}{\sqrt{2}} \left[(ds)(\bar d \bar s)+(su)(\bar s\bar u)\right].
\nonumber
\end{eqnarray}
Then one sees that the strange-quark counting successfully predicts the observed mass pattern:
$m(\sigma) < m(a_0) \sim m(f_0)$.
It also explains why the $J=0$ state is isolated from $J=1$ and $2$ partners and thus helps to explain anomalies in the scalar-meson nonets.
There are other hadrons which are pointed out to have possible exotic multi-quark components,
such as $D_s^*$, $X(3872)$ and $\Lambda(1405)$,
where the last one may be a baryon resonance composed of 5 quarks~\cite{Terasaki,Lipkin,Maiani,DeSwart,Strottman}.

Does the QCD dynamics allow such states?
We actually have a simple reasoning why the scalar mesons have significant 4-quark components.
As $q\bar{q}$ states, the scalar mesons, which have positive parity, must contain orbital excitation $L=1$ ($P$-wave mesons).
It is generally accepted that the orbital excitation in the quark model requires additional $500\> \mathrm{MeV}$, 
for instance, as is seen from the mass differences between the positive- and the negative-parity baryons; $N(1535)-N(940)$.
On the other hand, the 4-quark components, $qq\bar{q}\bar{q}$, may be realized in $L=0$ without orbital excitation 
and thus have an advantage over the $P$-wave excited states.
In fact, the cost of having an extra $q\bar{q}$ pair may be about $500~\> \mathrm{MeV}$;
the average mass of $\pi$ and $\rho$ mesons.
In terms of the diquarks, the combination of two scalar diquarks in $L=0$ is quite preferable in the standard quark-model 
and may easily reproduce the observed spectrum of the scalar mesons.
Therefore it is quite interesting and important to answer the question whether 
QCD indeed gives multi-quark states a lower energy and induces mixings of the different Fock components.

Our purpose in this paper is to study the possible mixing of different Fock components in the QCD-sum-rule approach.
It is important to establish a well-controlled formulation of the mixings of two (or more) types of operators in producing hadrons from the vacuum.
We also examine the definition of the Fock-state-mixing parameters 
so that we can compare the results with the predictions of the quark models.
In this paper, we focus on the 4-quark components of the flavor non-singlet scalar mesons, $a_0$ and $K_0^\ast$,
and demonstrate how our formulation works in this case.

In Sect.~\ref{sec:2}, we present the basic ideas of the QCD-sum-rule approach~\cite{SVZ,RRY} and 
give the interpolating field operators corresponding to the $q\bar q$ and $qq\bar q \bar q$ components of the scalar mesons.
In Sect.~\ref{sec:3}, we consider the definition of the mixing parameters of the two Fock components.
Two distinct methods of defining the mixing parameters are presented.
In Sect.~\ref{sec:4}, we give the results of the sum rules and study their significances.
The conclusion is given in Sect.~\ref{sec:5}.

\section{The QCD sum rules}\label{sec:2}
The sum rule is obtained by expressing the two-point function 
 \begin{equation}
  \Pi(p^2) = i\int d^4\! x \>e^{ipx}\langle 0|T[J(x)J^\dagger(0)]|0\rangle, \label{eq:2-1}
 \end{equation}
in two ways. One of them is based on the operator product expansion~(OPE),
where Eq.~(\ref{eq:2-1}) is calculated in deep Euclidean region ($-p^2\rightarrow\infty$)
and is described in terms of the QCD parameters, such as the quark condensate $\qq$, the gluon condensate $\GG$, the current-quark masses $m_q$ and so on.
The other one is based on a phenomenological parametrization of the spectral function.
The spectral function at the physical region ($p^2>0$)
is assumed to have a sharp-peak resonance at $p^2=m^2$ and continuum at $p^2>s_{\rm{th}}$:
 \begin{equation}
  \rho_{\mathrm{phen.}}(p^2)=|\lambda|^2\delta(p^2-m^2) +\theta(p^2-s_{th})\rho_{\mathrm{OPE}}(p^2). \label{eq:2-2}
 \end{equation}
The sum rule is obtained by matching the two expressions
so that the mass of the resonance, $m$, and the other phenomenological parameters can be determined from the QCD parameters.
The two expressions of the correlators are connected by a dispersion relation.
To simplify the sum rule, we approximate the continuum spectrum as the same form of the spectral function on the OPE side.

We further apply the Borel transformation, which is,
 \begin{equation}
  \mathcal{B}\equiv \lim_{-p^2,n^2\to \infty}\frac{(-p^2)^{n+1}}{n!}\left(\frac{d}{dp^2}\right)^n, \label{eq:2-3}
 \end{equation} 
where the limit is taken with $ M^2\equiv\frac{-p^2}{n^2}$ fixed.
It suppresses large-$p^2$ region by the factor of $e^{-p^2/M^2}$,
and thus the pole contribution is enhanced.
Due to the $p^2$ derivatives, the subtraction terms vanish.
The Borel transformation also suppresses the effects of the higher-dimensional terms in the OPE.

We employ the following local operators for $a_0^-$:
 \begin{eqnarray}
  J_2(x)&=&\left(\bar{u}_a(x)d_a(x)\right),\nonumber\\
  J_4(x)&=&\epsilon_{abc}\epsilon_{dec}\left(d^T_a(x)C\gamma_5s_b(x)\right)\left(\bar{s}_d(x)\gamma_5C\bar{u}^T_e(x)\right), \label{eq:2-4}\nonumber\\
 \end{eqnarray}
where $a,b,\ldots$, represent the color and $C=i\gamma^2\gamma^0$. 
The 2-quark operator, $J_2$, is uniquely determined,
while a specific 4-quark operator, $J_4$, is taken so as to consist of two $0^+$-diquark operators.
The flavor structure of $J_4$ is arranged to be $\bar{U}D$, and thus identical to $J_2$.
Similarly, the 2-quark and the 4-quark operators for $K_0^\ast$ are given by the $SU_f(3)$ rotation from the fields for $a_0$, \emph{i.e.}, $u\to d,d\to s,s\to u$ in Eq.~(\ref{eq:2-4}).

The number of independent 4-quark operators is discussed in a previous paper~\cite{Chen}.
There are two kinds of flavor-octet 4-quark operators:
i.e., $\bar{3}_c\times 3_c$ and $6_c\times\bar{6}_c$. Each kind allows five independent operators; 
thus altogether there are ten possible operators.

It is, however, difficult to find the most suitable operator for the QCD sum rule of the scalar mesons. 
In Ref.\cite{Chen}, the authors attempt to find the best combination.
It is interesting to employ such an operator for the present method,
while the computation may be much more complicated.

We here employ, instead, the simplest one composed of the $\bar{3}_f$ scalar diquarks,
$J^P=0^+$ and $C=\bar{3}$, which are supposed to be bound most strongly.

Our results of the OPE side are listed in Appendix \ref{ap:A}.
The OPE is truncated after the dimension-6 terms for $\Pi_{22}(p^2)$ (see Eq.~(\ref{eq:app1})).
In order to deal with the same order of power corrections,
we expand $\Pi_{24}(p^2)$ (Eq.~(\ref{eq:app2})) up to dim-9 and $\Pi_{44}(p^2)$ (Eq.~(\ref{eq:app3})) up to dim-12.
High dimensional operators, such as $\langle\bar{q}\bar{q}qq\rangle$ and $\langle\bar{q}\bar{q}qqG\rangle$ 
are evaluated by the vacuum-saturation approximation.

\begin{figure}[htb]
\includegraphics[width=0.45\linewidth]{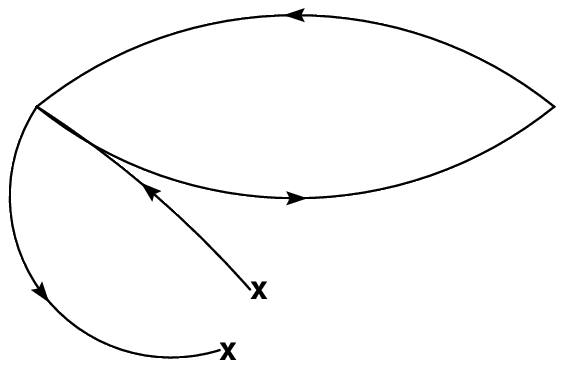}
\includegraphics[width=0.45\linewidth]{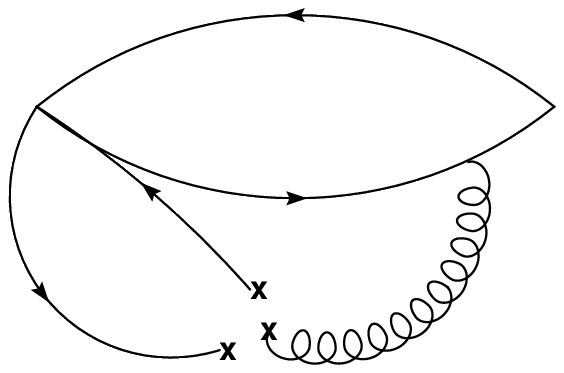}
\caption{The annihilation diagrams.}\label{fig:2-1}
\end{figure}

We consider the diagrams containing the $q\bar{q}$ annihilation
in order to take into account the mixing of different Fock states.
The quark-pair annihilations are substituted by $\qq$ or $\qGq$ (see FIG.\ref{fig:2-1}).
Because the interpolating fields (\ref{eq:2-4}) have the normal ordering,
the perturbative part of the $q\bar{q}$ annihilation must disappear.
Because the OPE is represented as a polynomial in $x$,
only the zero-th order term survives in the $x\to 0$ limit. 

\section{mixing of Fock states}\label{sec:3}
The main goals of this study are to determine
whether the scalar mesons are dominated by $q\bar q$ or $qq\bar q\bar q$, and if they are mixed,
to calculate their probabilities.
It turns out that such a mixing is not easily quantified.
It would be natural to consider the strengths of the couplings of the 2-quark and 4-quark operators to the physical state
and then evaluate the mixing angle.
However, such a procedure is strongly dependent on the definition and normalization of the local operators.
Indeed, a numerical factor can be easily hidden in the local operators 
and thus the magnitudes of the coupling strengths are ambiguous.

This problem happens to be more fundamental than just the definition of the operators.
In quantum mechanics, mixings of Fock states may occur
if the Hamiltonian contains terms which change the number of particles.
Then the wave function of an eigen-state can be written uniquely as a linear combination of the Fock components,
each of which is normalized properly.
In the field theory, however, the Fock-space separation may not be unique
as the number of quarks, which is defined by the number of quarks $+$ number of antiquarks, is not a conserved quantum number.
One may not be able to ``measure'' the number of quarks without ambiguity
because no conserved charge corresponding to the number of quarks is available.
Thus we have to consider the concept of the ``number of quarks''
in the context of the quantum-mechanical interpretation of the field-theoretical state.

Here, we propose two ways to define the ratio of the Fock-space probability 
by taking the mixing of $q\bar q$ and $qq \bar{q}\bar{q}$ in the scalar ($a_0$) meson as an example.

In the first approach, we define local operators ``normalized''
in the context of a full 4-quark operator $J_4$ in Eq.~(\ref{eq:2-4}).
In fact, the 4-quark operator $J_4$ contains $q\bar{q}$ component,
which is obtained by contracting the $\bar{q}q$ pair by the
quark condensate:
 \begin{equation}
  J_4(x)=J_4^\prime(x) + \underbrace{\frac{1}{6}\sss J_2(x)}_{J_2^\prime}. \label{eq:3-1}
 \end{equation}
We regard $J_2^\prime$ and $J_4^\prime$ as ``normalized'' 2-quark and 4-quark fields, respectively.
The quark condensate gives the dimension of normalization.
Using $J_2^\prime(x)$ and $J_4^\prime(x)$,
we define the mixing parameter, $\theta$, by
 \begin{eqnarray}
  \langle 0|J_2^\prime(x)|a_0\rangle &=& \lambda\cos\theta\>\phi(x),\nonumber\\
  \langle 0|J_4^\prime(x)|a_0\rangle &=& \lambda\sin\theta\>\phi(x).\label{eq:3-1.1}
 \end{eqnarray}
This definition happens to be equal to defining $\theta$ 
so that $J_a(x)=\cos \theta J_2^\prime(x) +\sin\theta J_4^\prime(x)$ couples to the physical state most strongly:
 \begin{eqnarray}
  \langle 0|J_a(x)|a_0\rangle &=& \lambda\phi(x),\label{eq:3-2}
 \end{eqnarray}
where $\phi(x)$ denotes the wave function of the center-of-mass motion of $a_0$
(i.e., a plane wave for a momentum eigenstate).

Then the mixing parameter can be evaluated from the following three correlation functions with an assumption 
that the poles are at the same position:
 \begin{eqnarray}
  \frac{1}{\pi}\mathrm{Im}\>i\!\int d^4\!x \>e^{ipx} \langle 0|T[J_2^\prime(x)(J_2^\prime)^\dagger(0)] |0\rangle & &\nonumber\\
  & &\hspace{-10em}=\delta(p^2-m^2)|\lambda|^2 \cos^2 \theta +\mathrm{cont.} , \nonumber\\
  \frac{1}{\pi}\mathrm{Im}\>i\!\int d^4\!x \>e^{ipx} \langle 0|T[J_4^\prime(x)(J_4^\prime)^\dagger(0)] |0\rangle & &\nonumber\\
  & &\hspace{-10em}=\delta(p^2-m^2)|\lambda|^2 \sin^2 \theta +\mathrm{cont.} , \nonumber\\
  \frac{1}{\pi}\mathrm{Im}\>i\!\int d^4\!x \>e^{ipx} \langle 0|T[J_2^\prime(x)(J_4^\prime)^\dagger(0)] |0\rangle & &\nonumber\\
  & &\hspace{-10em}=\delta(p^2-m^2)|\lambda|^2 \frac{\sin 2\theta}{2} +\mathrm{cont.} . \label{eq:3-3}
 \end{eqnarray}
This definition of the mixings is model independent, 
but it depends on the choice of the local operators.
Therefore, it does not necessarily have a direct relation to the mixing parameters employed in the quark models.

In order to define a mixing angle more appropriately to the quark models,
one may determine the normalization of the operators using quark-model wave functions.
For instance, the local operators in Eq.~(\ref{eq:2-4}), 
can be normalized to the wave functions of $q\bar q$ or $qq\bar q\bar q$ states in the MIT bag model:
 \begin{eqnarray}
  |(\bar{q}q)_{P_{0}}\rangle
&=&\frac{i}{\sqrt{2}}(|PS\rangle +|SP\rangle )\nonumber\\
&&\otimes\frac{1}{\sqrt{2}}\left(|\uparrow\downarrow\rangle-|\downarrow\uparrow\rangle\right)
\otimes
|\bar{u}d\rangle
\otimes\frac{1}{\sqrt{3}}\delta_{ab}|\bar{a}b\rangle,
\nonumber\\
|(qq{\bar q}{\bar q})_{S_{0}}\rangle
&=&|SSSS\rangle\nonumber\\
&&\otimes\frac{1}{2}\left(
|\uparrow\downarrow\uparrow\downarrow\rangle
-|\uparrow\downarrow\downarrow\uparrow\rangle
-|\downarrow\uparrow\uparrow\downarrow\rangle
+|\downarrow\uparrow\downarrow\uparrow\rangle
\right)\nonumber\\
&&
\otimes\frac{1}{2}\epsilon_{\alpha\beta u}\epsilon_{\delta\rho d}|\alpha\beta\bar{\delta}\bar{\rho}\rangle
\otimes\frac{1}{2\sqrt{3}}\epsilon_{abc}\epsilon_{dec}|ab\bar{d}\bar{e}\rangle,\nonumber\\
\label{eq:3-3.5}
\end{eqnarray}
where $\alpha,\beta,\ldots$ represent the flavor and $a,b,\ldots$ represent the color.
Then we compute the matrix elements
\begin{eqnarray}
\langle 0|J_2(0)|(\bar{q}q)_{P_{0}}\rangle&=&\lambda_2\phi(0),\nonumber\\
\langle 0|J_4(0)|(qq{\bar q}{\bar q})_{S_{0}}\rangle&=&\lambda_4\phi(0), \label{eq:3-4}
 \end{eqnarray}
where $\phi(0)$ denotes the center-of-mass wave function of the bag-model state.
Now, assuming the bag-model states (with definite number of quarks) are normalized properly,
one can use $\lambda_2$ and $\lambda_4$ for the normalizations of the operators $J_2$ and $J_4$, respectively.

In calculating the mixing parameter,
one needs only the ratio of $\lambda_2$ and $\lambda_4$, which is given by
 \begin{eqnarray}
 \frac{\lambda_4}{\lambda_2}\!\!&=&\!\!-\frac{i}{4\pi}\frac{\{{\cal N}_4(s_{1/2})\}^4}{{\cal N}_2(s_{1/2}){\cal N}_2(p_{1/2})}
\approx -0.24i\frac{R_2^3}{R_4^6},\label{eq:3-5}\\
 {\cal N}_n(S_{1/2})\!\!&=&\!\!\frac{ER_n}{R_n^{3/2}|j_0(ER_n)|}\frac{1}{\sqrt{2ER_n(-1+ER_n)}},\nonumber\\
 {\cal N}_n(P_{1/2})\!\!&=&\!\!\frac{ER_n}{R_n^{3/2}|j_1(ER_n)|}\frac{1}{\sqrt{2ER_n(1+ER_n)}},\nonumber\\
\textrm{and}\hspace{2em}ER_n\!\!&=&\!\!\left\{\begin{array}{cl}
           2.04&\qquad \textrm{for} \quad S_{1/2}\\
           3.81&\qquad \textrm{for} \quad P_{1/2}
          \end{array}\right.,\nonumber
 \end{eqnarray}
where $R$ is the bag radius and gives the dimensional scale of the normalizations
of the two operators that have different dimensions.
We here assume that the bag radius of the $q\bar{q}$ state is same as that of $qq\bar{q}\bar{q}$ state.
This assumption may be necessary to consider their mixing in the physical state.
It is, in fact, not a bad assumption because the bag radius of the $q\bar q$ state
must be larger, because of the relative $P$-wave motion, 
while that of the 4-quark state is large, because it has more quarks.

Thus, the physical state for $a_0$ is given by the mixings of the two states:
 \begin{eqnarray}
  |a_0\rangle&=& i\cos \theta |(\bar{q}q)_{P_{0}}\rangle + \sin\theta |(qq{\bar q}{\bar q})_{S_{0}}\rangle\nonumber\\
              &=& \cos \theta |a_0(2q)\rangle +\sin\theta |a_0(4q)\rangle. \label{eq:3-6}
 \end{eqnarray}
Note that here the factor $i$ is necessary to keep the phases of the 2- and 4-quark states
in accord so that the mixing parameter can be defined as a real parameter.

In the QCD sum rules,
the mixing parameter with the bag-model normalization can be calculated as
 \begin{equation}
  \frac{\mathrm{Im}\int\!d^4\! x\> e^{ipx}\langle 0|J^\prime_4(x)(J_4^\prime)^\dagger(0) |0\rangle}
  {\mathrm{Im}\int\!d^4\! x\> e^{ipx} \langle 0|J_2(x)J_2^\dagger(0) |0\rangle}
   =\left|\frac{\lambda_4}{\lambda_2}\right|^2\tan^2\theta. \label{eq:3-7}
 \end{equation}
This definition is model dependent, but it gives a direct interpretation associated with the quark model.

\section{results}\label{sec:4}
\begin{table*}
\caption{\label{tab:table1}
Standard values of the QCD parameters.}\label{tab:parameter}
\begin{ruledtabular}
\begin{tabular}{cccccc}
$m_s$ & $\qq$  & $\sss$ & $m_0^2 \equiv \qGq / \qq$ & $\GG$ &$\alpha_s$\\
\hline
$0.12~{\rm GeV}$ & $(-0.23~ {\rm GeV})^3$ & $0.8\times \qq$ & $0.8~{\rm GeV}^2$ & $(0.33~{\rm GeV})^4$ &0.4 \\
\end{tabular}
\end{ruledtabular}
\end{table*}
\begin{figure}[htb]
\includegraphics{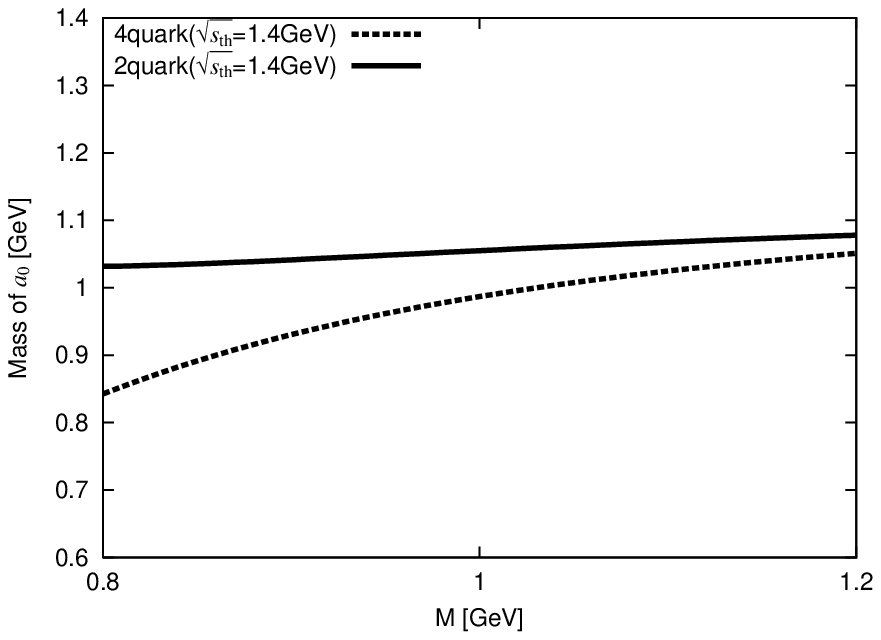}
\caption{The masses of $a_0$ for the pure 2-quark and pure 4-quark operators as functions of the Borel mass, $M$. The threshold parameter, $\sqrt{s_{\rm{th}}}$, is fixed to $1.4~\rm{GeV}$.}\label{fig:4-1}
\end{figure}
The mass formula is given by
 \begin{eqnarray}
  m=\sqrt{\frac{\frac{\partial}{\partial(-1/M^2)}\mathcal{B}\left[\Pi_{OPE}(p^2)\theta(s_{{\mathrm{th}}}-p^2)\right](M^2)}
{\mathcal{B}\left[\Pi_{OPE}(p^2)\theta(s_{{\mathrm{th}}}-p^2)\right](M^2)}}.
 \end{eqnarray}
We plot the mass of $a_0$ in the case of pure 2-quark and pure 4-quark in FIG.~\ref{fig:4-1}.
We use the values of the QCD parameters given in Table {\ref{tab:parameter}}, where $\uu\equiv \dd\equiv \qq$.
We neglect the masses of the $u$- and the $d$-quarks.
We see that the Borel stability is fairly good.
The positions of the poles of the 2-quark and the 4-quark are close to each other.
The 4-quark state is slightly lighter.

\begin{figure}[htb]
\includegraphics{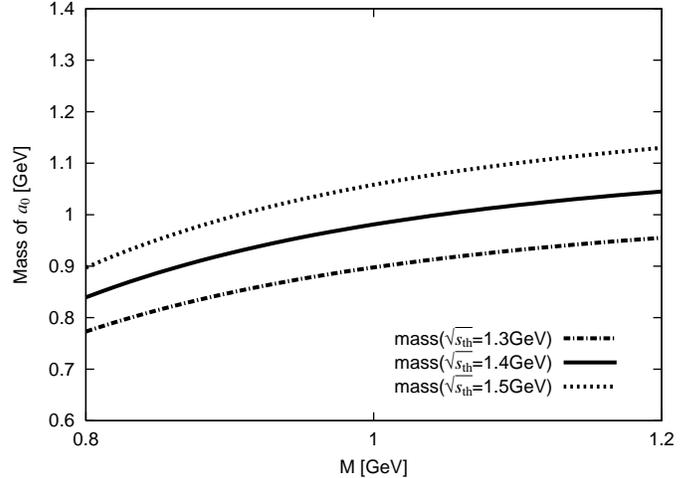}
\caption{Masses of $a_0$ for three choices of the threshold parameter, $\sqrt{s_{\rm{th}}}=1.3, 1.4$ and $1.5~\rm{GeV}$.
The 2-quark and 4-quark operators are mixed.
}\label{fig:4-2}
\end{figure}
We plot the mass of the mixed $a_0$
for various threshold parameter values, $s_{\rm{th}}$, in FIG.~\ref{fig:4-2}.
The predicted mass extracted from the mixed operator, $J_a(x)$, is about $0.9\sim 1.1~\mathrm{GeV}$,
which is similar to the result from the pure-4-quark operator given in FIG.~\ref{fig:4-1}.

\begin{figure}[htb]
\includegraphics{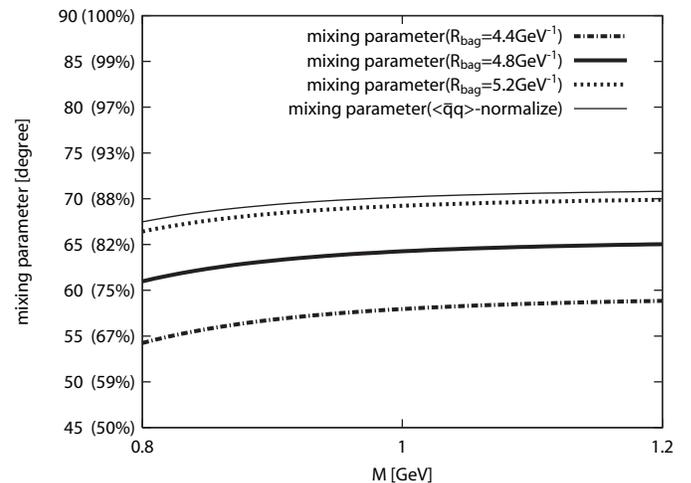}
\caption{The mixing parameters of $a_0$ plotted as functions of the Borel mass, $M$.
The threshold parameter, $\sqrt{s_{\rm{th}}}$, is fixed to $1.4~\rm{GeV}$.}\label{fig:4-3}
\end{figure}
The mixing parameters from the first-normalization method 
and the normalized operators according to the bag-model wave functions
are plotted for various bag radii in FIG.~\ref{fig:4-3}.
We observe that the mixing parameter is almost independent of the Borel mass, $M$.
We find that it is independent of threshold parameter, $s_{\rm{th}}$, too.
The weak $M$-dependence of the mixing parameter, which is dimensionless,
is attributed to the cancellation of the $p$-dependences ($M$-dependences) of the 2-quark and 4-quark correlators.
The mixing parameter from the first-normalization method is about $70$ degrees; in other words, the 4-quark component occupies $90\%$ of $a_0$.
In the bag model, we take the central value as $R=4.8~\mathrm{GeV}^{-1}$, which is the bag radius determined for the $\bar{q}q$ component of $a_0$~\cite{Jaffebag}.
The mixing parameter for $R=5.2~\mathrm{GeV}^{-1}$ is almost the same as the one in the first-normalization method.
We see that the dominant component of $a_0$ (about more than $70\%$) is the 4-quark state for  $R>4.4\mathrm{GeV}^{-1}$ and for the first-normalization method.

\begin{figure}[htb]
\includegraphics{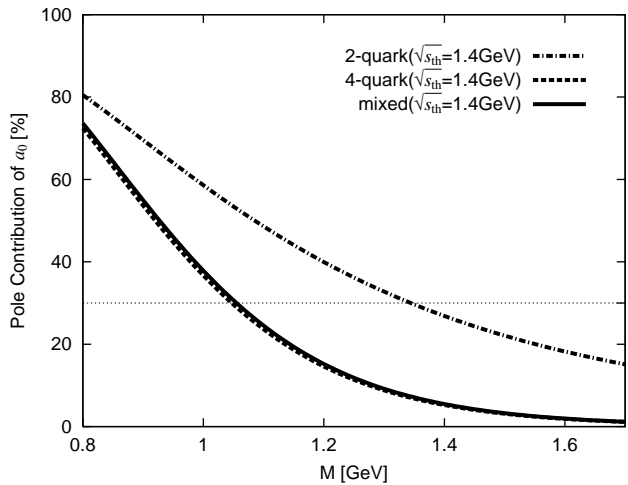}
\caption{The pole contribution defined by Eq.(\ref{eq:4-1}) for $a_0$ is plotted as a function of the Borel mass, $M$, in the cases of pure 2-quark components, pure 4-quark components and a mixing between them.
The threshold parameter, $\sqrt{s_{\rm{th}}}$, is fixed to $1.4~\rm{GeV}$.}\label{fig:4-4}
\end{figure}
We examine how well these sum rules work.
We plot the pole contribution defined by
\begin{equation}
 \frac{\mathcal{B}\left[\Pi_{\mathrm{OPE}}(p^2)\theta(s_{th}-p^2)\right]}{\mathcal{B}\left[\Pi_{\mathrm{OPE}}(p^2)\right]}, \label{eq:4-1}
\end{equation}
in FIG.~\ref{fig:4-4}.
The valid Borel window is taken as the region where the pole contribution is more than $30\%$.
This constraint is weaker than the one usually adopted \cite{RRY}, 
but we observe that the results are not sensitive to the choice of this value.
In FIG.~\ref{fig:4-4}, it is seen that the pole contribution for the pure 4-quark correlator is less than the one for the pure 2-quark correlator.
The reason is that the OPE of the 4-quark correlator contains higher powers of $p^2$ and grows rapidly for large $p^2$.
The sum rule for the mixed operator shows similar results.
We set the Borel window $M~<1.2\mathrm{GeV}$.
We observe form FIG.\ref{fig:4-1} that the positions of the poles of the 2-quark and 4-quark correlators begin to separate in this region.
If the positions of the poles are too different,
then the mixing will disappear.
But in this case the results for the mixed operator and the 4-quark operator are smoothly connected.
We conclude that the 4-quark component is dominant in $a_0$.
\begin{figure}[tb]
\includegraphics{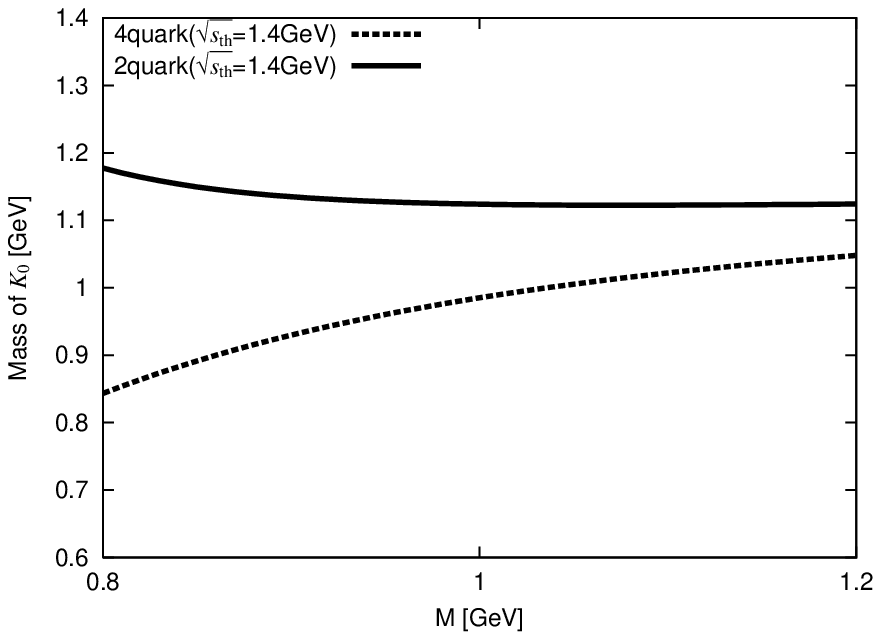}
\caption{The masses of $K_0^\ast$ for the pure 2-quark and pure 4-quark operators as functions of the Borel mass, $M$. The threshold parameter, $\sqrt{s_{\rm{th}}}$, is fixed to $1.4~\rm{GeV}$.}\label{fig:4-5}
\end{figure}
The results of $K_0^\ast$ for pure 2-quark and 4-quark components are presented in FIG.~\ref{fig:4-5}.
As compared to the Borel behavior of $a_0$ in FIG.~\ref{fig:4-1},
the mass extracted from the 2-quark correlator for $K_0^\ast$ is heavier,
while that of the 4-quark components lies at about the same mass as $a_0$.
The experimental value of the $K_0^\ast$ mass has been reported as about $0.84~\mathrm{GeV}$,
and it seems to have a large width: about $600~\mathrm{MeV}$~\cite{PDG}.
The scalar-nonet spectrum from our sum rule with the 4-quark component shows better agreement with the observed values than that with the 2-quark operator.
\begin{figure}[tb]
\includegraphics{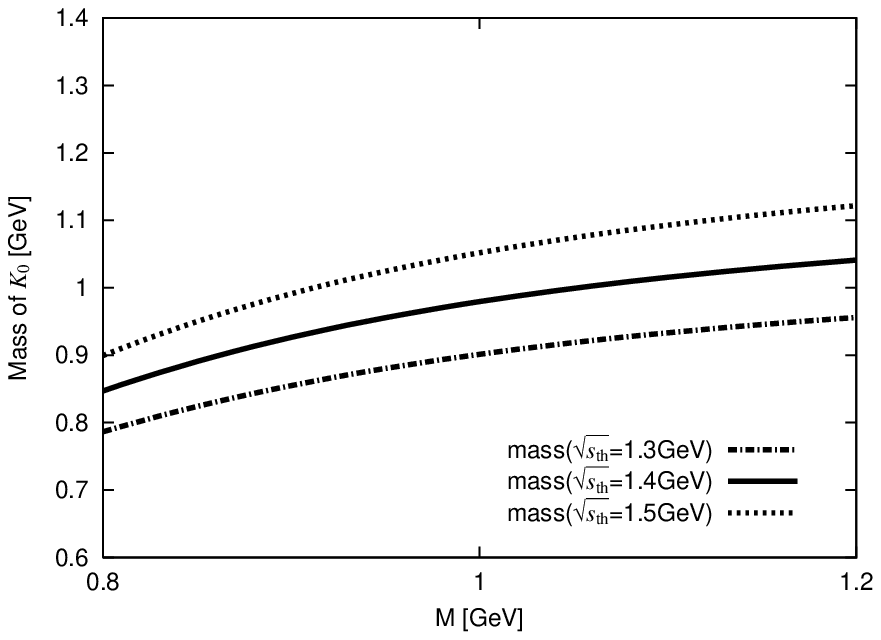}
\caption{Masses of $K_0^\ast$ for three choices of the threshold parameter, $\sqrt{s_{\rm{th}}}=1.3, 1.4$ and $1.5~\rm{GeV}$.
The 2-quark and 4-quark operators are mixed.}\label{fig:4-6}
\end{figure}
\begin{figure}[tb]
\includegraphics{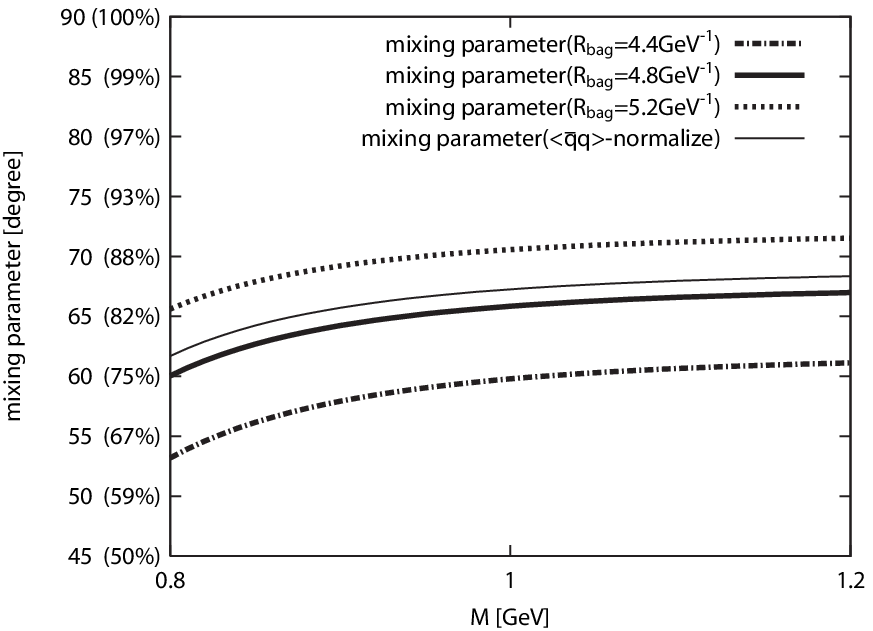}
\caption{The mixing parameters of $K_0^\ast$ plotted as functions of the Borel mass, $M$.
The threshold parameter, $\sqrt{s_{\rm{th}}}$, is fixed to $1.4~\rm{GeV}$.}\label{fig:4-7}
\end{figure}
The position of the pole of $K_0^\ast$ for the 2-quark operators starts splitting at low $M$ values from that of the 4-quark operators.
We, however, attempt to evaluate the mixing parameter by assuming that the positions of the poles are at the same position.
The results are shown in FIG.~\ref{fig:4-6} and FIG.~\ref{fig:4-7},
which correspond to FIG.~\ref{fig:4-2} and FIG.~\ref{fig:4-3} in the case of $a_0$, respectively.
One sees that the 4-quark component is dominant, occupying about $70\%\sim 90\%$ of $K_0^\ast$.
These results for $K_0^\ast$ are similar to those for $a_0$.

\begin{figure}[tb]
\includegraphics{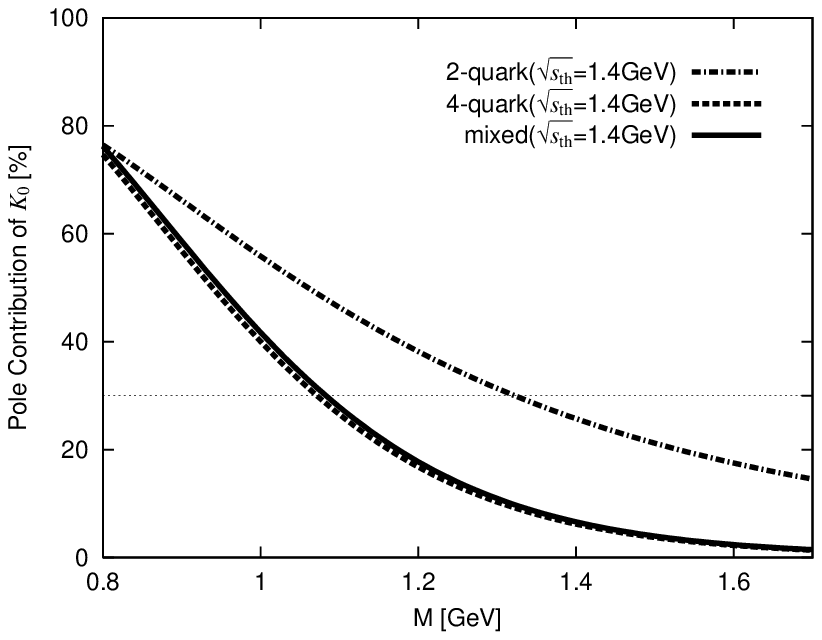}
\caption{The pole contribution defined by Eq.(\ref{eq:4-1}) for $K_0^\ast$ is plotted as a function of the Borel mass, $M$, in the cases of pure 2-quark components, pure 4-quark components and a mixing between them.
The threshold parameter, $\sqrt{s_{\rm{th}}}$, is fixed to $1.4~\rm{GeV}$.}\label{fig:4-8}
\end{figure}
The pole contribution for $K_0^\ast$ is shown in FIG.~\ref{fig:4-8}.
We observe from FIG.~\ref{fig:4-4} and FIG.~\ref{fig:4-8} that the behavior is similar to that for $a_0$.

\section{conclusion}\label{sec:5}
A formulation is proposed to take into account the mixing between different Fock states in the QCD sum rules.
In order to quantify the mixing probability,
one needs the normalization of the operators.
We suggest two ways:
One is to define the ``normalization" using a multiquark operator which couples to the ground state most strongly.
The other way is to adjust the operators to the normalized wave functions from the MIT bag model.

We apply the formulation to $a_0$ and $K_0^\ast$ which are members of the scalar nonet.
Our sum rules indicate that $70\sim 90\%$ of these scalar mesons is composed of the 4-quark components.
We find that $K_0^*$ is almost degenerate with $a_0$.

There exist several other studies of the 4-quark scalar nonet in the QCD sum rule.
Two of them~\cite{Chen,Wang} are consistent with our work in reproducing the lighter 4-quark mass than 2-quark mass.
In these analyses, the $K_0^\ast$ is predicted to have a smaller mass than $a_0$.
It should, however, be noticed that the predicted masses depend on the threshold parameter rather strongly.
Therefore, one may choose the threshold
so that $K_0^*$ is lighter than $a_0$ reproducing experimental data.
$K_0^*$ is a broad resonance which is not completely established.
Therefore we do not further examine the mass difference of $a_0$ and $K_0^*$.
Another work~\cite{Lee} claims that no signal is obtained for the 4-quark scalar nonet.
We note that their analysis differs from ours 
in not considering the mixing diagrams, truncating dimension in the OPE, interpolating field and
the definition of the coupling strength of ground state.

It is extremely interesting that some of the excited hadrons are accounted for by considering exotic multi-quark components.
There are several other ``anomalous'' hadrons which may contain exotic multi-quark components.
Among them, the P-wave baryons, in particular $\Lambda(1405)$, are strong candidates to have 5-quark components.
In another paper\cite{Nakamura},
we report the results of applying the current method of the Fock-space mixing to the flavor-singlet $\Lambda$ state.
Multi-quark mesons may also be found in heavy quark systems,
where newly found states do not fit well in the $q\bar{q}$ spectrum and thus are suspected to have 4-quark components.

It is also important to examine the decay widths and the branching ratios of those multi-quark states.
The mechanism for the fall-apart decay in which the multi-quark hadrons dissociate into two color-singlet hadrons without creating $q\bar{q}$ pairs.
The widths associated with the fall-apart processes depend strongly on the quantum numbers as well as on the configurations of the multi-quarks.
As QCD does not forbid such multi-quark states, the width is the key to understand why we do not see many ``exotic'' hadrons in nature.
Their possibility should be further pursued both experimentally and theoretically.
\acknowledgements
The authors acknowledge Prof. A. Hosaka for discussions.
This work is partially supported by Grant for Scientific Research
((B) No. 15340072 and Priority Area No. 17070002) from the Ministry of Education, Culture, Science and Technology, Japan.
J.S. acknowledges the support of the Japan Society for the Promotion of Science (JSPS).
\appendix
\begin{widetext}
\section{the result of OPE}\label{ap:A}
The results of the OPE are summarized as:
\begin{eqnarray}
\Pi_{22}(p^2) &=& i\int d^4\! x \>e^{ipx}\langle 0|T[J_2(x)J_2^\dagger(0)]|0\rangle\nonumber\\
&=&-\frac{3}{8\pi^2}p^2\ln(-p^2) -\frac{1}{2p^2}(m_u\uu +m_d\dd )
-\frac{1}{p^2}(m_d\uu +m_u\dd )\nonumber\\
& &-\frac{1}{8p^2}\GG-\frac{16\pi\alpha_s}{27(p^2)^2}\left(\uu^2+\dd^2\right)-\frac{16\pi\alpha_s}{3(p^2)^2}\uu\dd\nonumber\\
& &+\frac{1}{2(p^2)^2}(m_u\dGd +m_d\uGu),\label{eq:app1}\\
\Pi_{24}(p^2)&=&i\int d^4\! x \>e^{ipx}\langle 0|T[J_2(x)J_4^\dagger(0)]|0\rangle
=i\int d^4\! x \>e^{ipx}\langle 0|T[J_4(x)J_2^\dagger(0)]|0\rangle\nonumber\\
&=& -\frac{1}{16\pi^2 }p^2\ln(-p^2)\sss -\frac{1}{32\pi^2}\ln(-p^2)\sGs \nonumber\\
             & &-\frac{1}{48 p^2}\sss\GG-\frac{1}{12 p^2}\{m_u(\uu+2\dd)+m_d(2\uu+\dd)\}\sss \nonumber\\
             & &+\left\{-\frac{8\pi\alpha_s}{81(p^2)^2}\sss\left(\uu^2+\dd^2\right)
-\frac{8\pi\alpha_s}{9(p^2)^2}\sss\uu\dd\right\}\nonumber\\
             & &+\frac{4\pi\alpha_s}{81(p^2)^2}\left(\uu +\dd\right)\sss^2\nonumber\\ 
             & &+\frac{1}{12(p^2)^2}\sss (m_u\dGd +m_d\uGu) +\frac{1}{24(p^2)^2}(m_u\dd +m_d\uu)\sGs,\label{eq:app2}\\
\Pi_{44}(p^2)&=&i\int d^4\! x \>e^{ipx}\langle 0|T[J_4(x)J_4^\dagger(0)]|0\rangle\nonumber\\
&=&-\frac{12}{4!\> 5!\> 2^8\pi^6}(p^2)^4\ln(-p^2)\nonumber\\
             & &-\frac{1}{3!\> 2^6\pi^4}(p^2)^2\ln(-p^2)\left\{m_u(\uu-2\sss)+m_d(\dd-2\sss)-2m_s(\uu+\dd-\sss)\right\}\nonumber\\
             & &-\frac{1}{3!\> 2^8\pi^4}(p^2)^2\ln(-p^2)\GG \nonumber\\
             & &+\frac{1}{96\pi^2}p^2\ln(-p^2)\sss(-\sss-4\uu-4\dd)\nonumber\\
             & &-\frac{1}{768\pi^4}p^2\ln(-p^2)(m_u\uGu+m_d\dGd+2m_s\sGs)\nonumber\\
             & &-\frac{1}{96\pi^4}p^2\ln(-p^2)(m_d\sGs +m_s\dGd +m_s\uGu +m_u\sGs)\nonumber\\
             & &+\frac{1}{192\pi^2}\ln(-p^2)\left\{[\uu+\dd]\sGs +\sss[\uGu+\dGd-2\sGs]\right\}\nonumber\\
             & &-\frac{1}{384 \pi^2}\ln(-p^2)(m_u\uu +m_d\dd +2m_s\sss)\GG\nonumber\\
             & &+\frac{1}{144 \pi^2}\ln(-p^2)(m_d\sss +m_s\dd +m_s\uu +m_u\sss)\GG\nonumber\\
             & &-\frac{1}{18p^2}\{ (m_u+m_d)\uu\dd\sss + m_s\sss^2 (\uu +\dd)\}\nonumber\\
             & &+\frac{1}{9p^2}\{ (m_u\dd+m_d\uu)\sss^2 + 2m_s\uu\dd\sss\}\nonumber\\
             & &-\frac{1}{24p^2}(m_u\uu+m_d\dd)\sss^2\nonumber\\
             & &-\frac{7}{864 p^2}(\uu+\dd)\sss\GG-\frac{1}{288p^2}\sss^2\GG\nonumber\\
             & &-\frac{41}{2^{10}\cdot 3^2\pi^2p^2}(\uGu+\dGd)\sGs\nonumber\\
             & &-\frac{1}{2^9\cdot 3\pi^2 p^2}\sGs^2\nonumber\\
             & &+\frac{20\pi\alpha_s}{27(p^2)^2}\uu\dd\sss^2-\frac{4\pi\alpha_s}{243(p^2)^2}\sss^2\left(\uu^2+\dd^2\right)\nonumber\\
             & &-\frac{8\pi\alpha_s}{243(p^2)^2}\sss\left(\sss^2\uu+\dd^2\uu+\sss^2\dd+\uu^2\dd\right)\nonumber\\
             & &+\frac{8\pi\alpha_s}{243(p^2)^2}\sss^3\left(\uu+\dd\right)\nonumber\\
             & &+\frac{1}{72(p^2)^2}\sss^2(m_u\dGd +m_d\uGu)\nonumber\\
             & &-\frac{1}{36(p^2)^2}\Biggl\{ \uu\sss(m_s\dGd+m_d\sGs)+\dd\sss(m_u\sGs+m_s\uGu)\Biggr\}\nonumber\\
             & &+\frac{1}{72(p^2)^2}\sGs\sss\left( m_u\dd +m_d\uu\right).\label{eq:app3}
\end{eqnarray}
\end{widetext}

\end{document}